\documentclass[twocolumn,twoside,slac_two,a4paper]{revtex4}
\usepackage{graphicx}
\usepackage{fancyhdr}
\newcommand{\nhi}{N(\mathrm{H\,\scriptstyle{I}})}

\newcommand{\nhd}{N(\mathrm{H_2})}
\newcommand{\wco}{W_\mathrm{CO}}
\newcommand{\hi}{\mathrm{H\,\scriptstyle{I}}}

\newcommand{\hd}{\mathrm{H}_2}
\newcommand{\xco}{X_\mathrm{CO}}
\newcommand{\ebv}{\mathrm{E(B-V)}}

\newcommand{\qhinull}{q_\mathrm{H\,\scriptscriptstyle{I}}}

\newcommand{\qconull}{q_\mathrm{CO}}
\newcommand{\qebv}{q_\mathrm{EBV}}

\newcommand{\g}{$\gamma$}
\newcommand{\mod}{\texttt{54\_71Xvarh7S}}

\newcommand{\apj}{Astrophys. J. }
\newcommand{\apjl}{Astrophys. J. Lett. }
\newcommand{\apjs}{Astrophys. J. Suppl. }
\newcommand{\aap}{Astron. \& Astrophys. }
\newcommand{\astropp}{Astroparticle Phys. }

\newcommand{\prl}{Phys. Rev. Lett. }
\newcommand{\pasj}{Publ. Astron. Soc. Jpn. }

\begin{document}

\title{The cosmic-ray puzzle and the census of the interstellar medium: the
\textit{Fermi} LAT view of Cassiopeia, Cepheus and the Perseus arm.}

\author{L. Tibaldo$^{1,2}$ and I. A. Grenier$^2$\\on behalf of the
\textit{Fermi} LAT collaboration}
\affiliation{$^1$ INFN -- Sezione Di Padova \& Dipartimento di Fisica
``G. Galilei'' -- Universit\`a di Padova,\\ I-35131 Padova, Italy
\newline
$^2$ Laboratoire AIM, CEA-IRFU/CNRS/Universit\'e Paris Diderot,
Service
d'Astrophysique, CEA Saclay, 91191 Gif sur Yvette, France}

\begin{abstract}
Diffuse \g{}-ray emission arising from interactions of cosmic
rays with the interstellar gas traces
the densities of
both of them throughout the Milky Way. 
We discuss the results obtained from the
analysis of \textit{Fermi}
LAT observations in the region of Cassiopeia and Cepheus, towards the
Perseus spiral arm. We find that the \g-ray emissivity of local
gas is
consistent with expectations based on the cosmic-ray spectra measured at the
Earth. The emissivity decreases from the Gould Belt to the Perseus arm, but the
measured gradient is flatter than predictions by a propagation model based on
a cosmic-ray source distribution peaking in the inner Galaxy as
suggested by pulsars. The $\xco=\nhd/\wco$ conversion factor moderately
increases by a factor $\sim 2$ from the Gould Belt to the Perseus arm. The
presence of additional gas not properly traced by $\hi$ and CO surveys in the
Gould Belt
is suggested by the correlation between \g-rays and thermal emission from cold
interstellar dust.
\end{abstract}

\maketitle

\thispagestyle{fancy}

\section{INTRODUCTION}
More than a century after the discovery of cosmic rays (CRs), their origin is still
mysterious and their propagation in the interstellar space subject of
long-standing debates. Interactions between CRs and interstellar gas give
rise to \g{} rays through $\pi^0$ decay and
Bremsstrahlung radiation. In addition, CR electrons and positrons produce
\g{} rays through inverse Compton scattering on the low energy interstellar
radiation field. Thus, \g{} rays can probe CR densities in the solar
neighborhood, 
beyond the direct measurements in the solar system, as well as in remote parts
of the Galaxy. 

On the other hand, the interstellar \g-ray emission usefully complements gas and
dust tracers at other wavelengths to probe the interstellar medium (ISM)
column densities.  Two main radio/microwave tracers of the
ISM have long been used in \g-ray astrophysics: the 21 cm line of atomic
hydrogen, $\hi$, and the 2.6 mm line of carbon monoxide, CO. There are, however,
open issues in the derivation of gas
column densities from these observations, and hints that their
combination does not
provide an exhaustive census of the gas in the ISM.

In this contribution, we add a few pieces to this intriguing puzzle with the use of
\textit{Fermi} Large Area Telescope (LAT) data for the second Galactic quadrant,
at $100^\circ \leq l \leq
145^\circ$, $-15^\circ \leq b \leq 30^\circ$. This region has been selected due
to
the presence of conspicuous atomic and molecular
complexes and the good kinematic separation between the different structures
seen along the line of
sight. Moving toward the outer Galaxy, the region encompasses complexes that are
part of the Gould Belt (within $\sim300$~pc from the
Solar System), of the local arm ($\sim 1$~kpc), of the Perseus arm ($2.5 -
4$~kpc) and of the outer spiral arm ($>5$~kpc). The
analysis is described in detail in~ \cite{cascep}. We summarize below the main
results and discuss their physical implications.

\section{GAMMA-RAY EMISSION FROM THE LOCAL INTERSTELLAR MEDIUM}
The \g-ray emissivity of atomic gas can be estimated by comparing \g-ray and radio
data because the 21 cm line of the $\hi$ hyperfine transition allows us to
derive hydrogen column densities, $\nhi$. The results presented
here are based on the often-used assumption (in \g-ray astrophysics) of a
uniform $\hi$ spin temperature of 125 K. There are, however, large
uncertainties in the gas spin temperature, as well as non-uniformities within a
single interstellar complex. The issue of the $\nhi$ determination will be
further discussed in Section~\ref{ISMtrac}.

\begin{figure}[!hbt]
\includegraphics[width=0.5\textwidth]{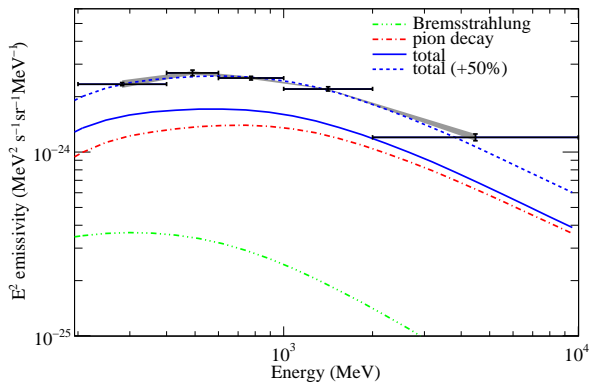}
\caption{Differential emissivity per $\hi$
atom measured in the Gould Belt. Black points are our measurements, the shaded
area gives the
systematic
uncertainties due to the LAT event selection efficiency. The lines represent the
predictions by the GALPROP model
\mod.}\label{locemiss}
\end{figure}

In Fig.~\ref{locemiss} we show the differential emissivity per $\hi$
atom measured in the Gould Belt.
The results are compared with the predictions by GALPROP, a widely used code
for CR propagation
 \cite{strong1998,strong2007}.
The results are consistent with an independent
study of LAT measurements at intermediate Galactic latitudes
in the third quadrant \cite{lochiemiss}. The $\hi$ emissivity in the
Gould Belt obviously exceeds the GALPROP prediction by $\sim 50\%$. 

However, the normalization of the spectrum is consistent with the
expectations within the actual constraints. In the energy range considered (200
MeV -- 10 GeV) the
emission is dominated by the pionic
component due to hadronic interactions. An important source of uncertainty
comes from systematic errors in CR measurements:
the current literature reports differences up to $\sim 20\%$ for the proton
spectrum. An additional $\sim 30\%$ uncertainty is due to the contribution from
heavier nuclei in both CRs and the ISM. The GALPROP model we adopted, following
the method by Dermer \cite{dermer1986a,dermer1986b}, predicts an
effective enhancement of $\sim 1.45$ with respect to pure $p$-$p$ emissivities.
Different authors, however, report values as high as
$1.75-2$~\cite{mori2009}\footnote{The origin of the discrepancies is still
unclear, either due to the calculation method (e.g. parametrization of the
reaction yields versus numerical simulations) and/or to differences in the CR
spectra within the experimental constraints.}.

Provided a 50\% scaling of the emissivity, the spectral shape is in good
agreement with the expectations. Thus we can conclude that CR nuclei in
the local ISM are similar to those measured at Earth: further constraints will
be given by improved direct measurements of CRs (by missions like PAMELA and
AMS02).

This is an additional evidence that LAT
measurements are not consistent with the GeV excess seen by EGRET
\cite[see e.g.][]{hunter1997,strong2004}, as already deduced by LAT observations
at
intermediate Galactic latitudes \cite{nongevexc}. A deviation with respect to
the expected spectral shape of diffuse \g-ray emission was
observed by EGRET over the whole sky for
energies $\gtrsim 1$ GeV,
leading to several possible interpretations: instrumental effects,
differences between typical CR spectra in the ISM and those in the proximities
of the Earth, dark matter annihilation. The GeV excess was noticed also
in the $\hi$ emissivity of local interstellar complexes \cite{digel2001}, being
inconsistent with the results of our analysis. The same kind of difference is
noticed in the spectra of bright point sources, notably for the Vela
pulsar~\cite{vela1}, and so the differences between the spectra measured by
EGRET and those measured by the LAT are plausibly instrumental in origin.

\section{THE $\protect{\mathbf{\xco}}$ GRADIENT IN THE OUTER
GALAXY}\label{xcograd}
The molecular phase of the ISM is difficult to trace, because its major
constituent, $\hd$, does not have observable emission lines in its cold phase.
The 2.6 mm line of the CO $J=1 \rightarrow 0$ transition has often been used as
a surrogate tracer of molecular masses. Observations, notably virial mass
calculations,
suggest that the velocity-integrated antenna temperature $\wco$, despite the
optical thickness, is proportional
to the total mass in the emitting region \cite{solomon1991}.

The conversion factor which transforms $\wco$ intensity into $\nhd$ column
density is known as $\xco=\nhd/\wco$. For many years it has been considered as uniform
across the Galaxy, but we have now evidence  from virial masses
\cite{digel1990}, from COBE/DIRBE observations \cite{sodroski1995}, and from the
$\xco$ dependence on metallicity in external galaxies
\cite{israel1997} that it increases with Galactocentric radius.
Such an increase was shown to be consistent with EGRET data
\cite{strong2004grad}, but the limited performance of previous
\g-ray telescopes did not allow to get accurate measurements in
the outer Galaxy \cite{digel1996,digel2001}.

High-energy \g-ray observations provide important constraints on $\xco$ values.
Since
the molecular binding energy is negligible with respect to the energy scale of
the \g{} radiation processes, the emissivity per $\hd$ molecule is twice the
emissivity per $\hi$ atom for gas threaded by the same CR flux. This statement
relies on the assumption that CR diffusion lengths are large enough to ensure
similar CR spectra in the atomic and molecular phases of a complex
(i.e. they do not vary significantly over distances of tens of pc) and
that CRs penetrate molecular clouds uniformly to their core in spite of the
enhanced magnetic fields (idea still under debate).

\begin{figure}[!hbt]
\includegraphics[width=0.5\textwidth]{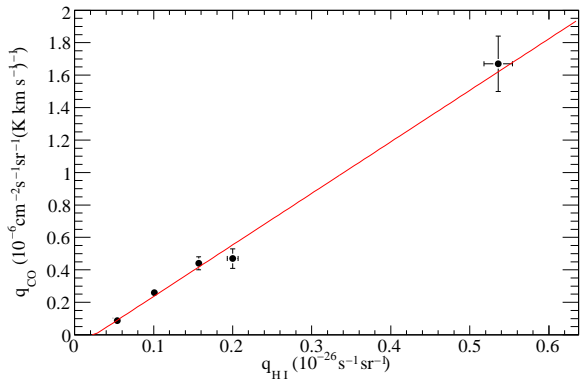}
\caption{Correlation between $\qconull$ and
$\qhinull$ over 5 energy ranges between 200 MeV and 10 GeV in the local arm
clouds. The solid red line represents the best linear fit. The fit uses the measurement uncertainties on both axes.}\label{qhico}
\end{figure}

The results of our work allowed us to correlate the integrated emissivity per
$\wco$ unit ($\qconull$) and per $\hi$ atom
($\qhinull$) in different regions along the line of sight, for several energy
bands from 200 MeV to 10 GeV. The results are shown for the local arm in
Fig.~\ref{qhico}.

\begin{figure}[!hbt]
\includegraphics[width=0.5\textwidth]{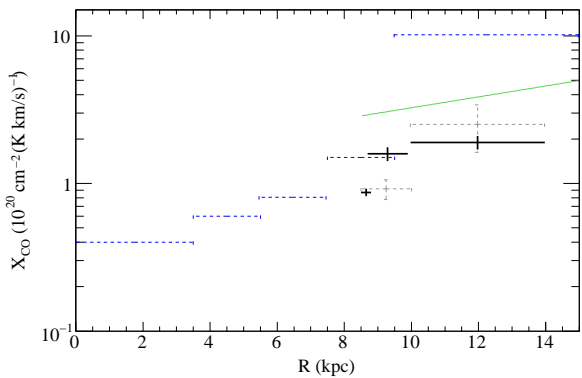}
\caption{Black points: $\xco$ values measured in the Gould belt,
local and Perseus arm (from left to right). Blue dashed line: model
by Strong et al.~\cite{strong2004grad}. Green solid line: the conversion
function based on
virial masses by Nakanishi and Sofue \cite{nakanishi2006}.}\label{xcor}
\end{figure}

The good linear correlation supports the assumption of similar spectra in
the atomic and molecular phases of the clouds and the energy-independent
penetration of CRs to the molecular cores. The slope of the
linear relation provides an estimate of the $\xco$ ratio. Good linear correlations are also 
found for the Gould Belt clouds and in the Perseus arm. The values of $\xco$ in
the different regions are summarized in
Fig.~\ref{xcor} as a function of Galactocentric radius.

The results confirm a significant but moderate increase of $\xco$
with Galactocentric radius. The gradient is consistent with the metallicity decrease in the outer Milky Way and the strong relation seen between
metallicity and $\xco$ in external galaxies \cite{israel1997}. This relation is
plausibly due to poor self-shielding against CO photodissociation by UV radiation in low metallicity environments.

Our results do not support the extremely large gradient proposed by Strong et al.~\cite{strong2004grad} beyond the
solar circle (dashed blue line in
Fig.~\ref{xcor}). The latter was introduced on the basis of non-\g-ray data to
alleviate the \emph{CR gradient problem} (see below Section~\ref{CRsec}).
We find a value significantly lower than the prediction of this model for the
molecular complex associated with NGC~7538 and Cas~A in the Perseus arm,
which is the most massive molecular complex beyond the solar circle.

The values of $\xco$ we measured are systematically lower than the conversion
law derived by Nakanishi and Sofue~\cite{nakanishi2006} from virial masses. We
have verified that a difference of comparable magnitude is found for our
clouds between the masses derived from the $\xco$ ratios just measured and the
virial masses (based on the same CO data used for the \g{}-ray analysis)
\cite{cascep}. An instrumental origin of the
discrepancy (e.g. calibration differences between different CO datasets) is
therefore unlikely. As already said, the \g-ray masses may be
biased by the non-uniform penetration of CRs within the dense cores of the
clouds. The limited spatial resolution of the \g-ray data may also lead to
separation problems between the molecular phase of a cloud and its dense atomic
envelope. On the other hand, virial masses rely on rather crude hypotheses:
spherical clouds with simple density profiles and turbulent support to hold a
cloud in equilibrium against gravitational collapse. The effects of magnetic
support are rarely considered (they were not in our calculation) and intrinsic
velocity gradients can broaden the velocity dispersion. The virial estimate also
corresponds to the total dynamical mass, whereas the \g-ray estimate corresponds
to the mass spatially associated with CO. Gas not properly accounted
for by $\wco$ intensities could explain, at least in part, the differences
between the virial and \g-ray masses (see below Section~\ref{ISMtrac}).

\section{THE COSMIC-RAY DENSITY GRADIENT IN THE
OUTER GALAXY}\label{CRsec}
The CR origin is still unclear. The EGRET non-detection of the Small
Magellanic Cloud \cite{lin1996}, an external galaxy member of the local group,
highlighted that its CR content significantly differs from
that of the Milky Way. The EGRET upper limit was recognized as evidence that
CRs,
at least below
$\sim 10^{15}$ eV, have a Galactic origin. Supernova
remnants (SNRs) have been considered for many years the best candidates as CR
accelerators in our Galaxy. Whereas diffusive
acceleration of electrons by SNR shock waves is obvious from multiwavelength
observations, only recently we started accumulating indirect evidence for the
acceleration of CR nuclei \g-ray
observations in the TeV
\cite{albert2007ic443,aharonian2008w28,acciari2009ic443} and GeV domain
\cite{lat2009w51}.

On the other hand, the distribution of SNRs in the Galaxy is very poorly
determined \cite{case1998} because of the sparsity of the known sample (a few tens)
and its large selection effects. Since the COS-B era it is well known that the gas
\g-ray emissivity gradient across the outer Galaxy is much
flatter than expected from the decline in SNR detections
\cite{strong1988}.

Strong et al.~ \cite{strong2004grad} proposed to use the
observed pulsar distribution to trace CR sources because of their richer sample
\cite{lorimer2004}. Large uncertainties in the distance derivation from the
dispersion measurements can, however, easily bias their radial distribution. The
decline in pulsar counts in the outer Galaxy is even steeper than the SNR one,
strengthening the \emph{CR gradient problem}. Strong et
al.~\cite{strong2004grad} attempted to alleviate this problem by increasing the
amount of molecular gas in the outer Galaxy, invoking
a strong increase of $\xco$ beyond the solar circle. Yet, tuning the $\xco$
ratio does not explain the flat $\qhinull$ emissivity gradient recorded in the
ubiquitous and massive atomic phase  \cite{strong1988,strong1996}. As
discussed in
Section~\ref{xcograd} this solution is disfavored by LAT measurements.

\begin{figure}[!hbt]
\includegraphics[width=0.5\textwidth]{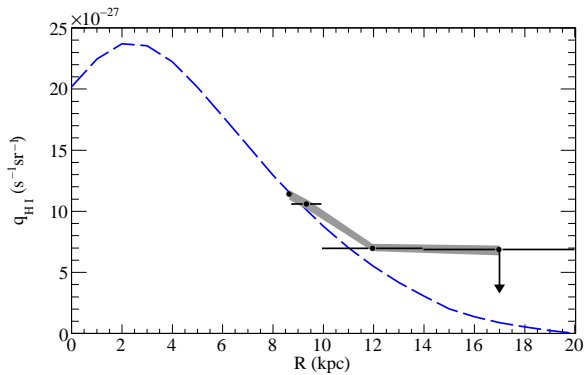}
\caption{$\hi$ emissivities  integrated
between 200 MeV and 10 GeV ($\qhinull$) measured in the Gould belt,
local, Perseus and outer arm (from left to right). Black points are our
measurements, the grey shaded area
represents the systematic uncertainties in the LAT event selection efficiency.
The
blue
dashed line gives the prediction by the GALPROP model \mod.}\label{qhifig}
\end{figure}

Our measurements of the $\hi$ emissivity across the outer Galaxy are shown in
Fig.~\ref{qhifig}.

The results are compared with the predictions by a GALPROP model which
uses a CR source distribution derived from pulsars. The \emph{CR gradient
problem} is confirmed by LAT data: the measured decline
in emissivity is much flatter than the prediction by GALPROP for
the pulsar (Fig.~\ref{qhifig}) and also the SNR distribution (\textit{Fermi}
LAT
collaboration, in preparation).

Systematically underestimating the remote $\nhi$ densities as the result of
self-absorption phenomena can lead to a bias in the emissivity
gradient.
Correcting for $\hi$ self absorption, Gibson et
al.~\cite{gibson2005} suggested that a
significant amount of $\hi$ mass has been overlooked by emission surveys in the
second quadrant for the outer Galaxy. The large emissivity in the outer Galaxy
may be explained also by a population of unresolved sources clustering in the
Perseus arm structures.

On the other hand, if the flat gradient is real, it challenges either the SNRs
as CR parent population or the modeling of CR diffusion across
the Galaxy. A possible explanation is provided by the large uncertainties on the
SNR distribution. Additionally, the CR propagation parameters are
usually derived from the isotopic composition of CRs measured at
Earth \cite[see e.g.][]{strong1998}, a
reasonable assumption, but not required by observational constraints.
An alternative scenario is given by non-uniform CR diffusion
\cite{evoli2008}.

\section{THE TRACERS OF THE INTERSTELLAR MEDIUM IN THE GOULD
BELT}\label{ISMtrac}
The modeling of the \g-ray interstellar emission based on the
combination of $\hi$ and CO
data is being challenged by observations. EGRET
data, compared with other tracers, already suggested that a significant amount
of gas is not properly traced by these radio/microwave lines
\cite{grenier2005}. The nearby, off-plane, clouds located in the Gould Belt are
well suited to probe ISM tracers, both because there is little confusion
along the line of sight and because the higher linear resolution of the \g-ray
data within the clouds allows a good separation of the emission arising from the
different phases and sources.

\begin{figure*}
 \begin{center}
  \begin{tabular}{cc}
\includegraphics[width=0.5\textwidth]{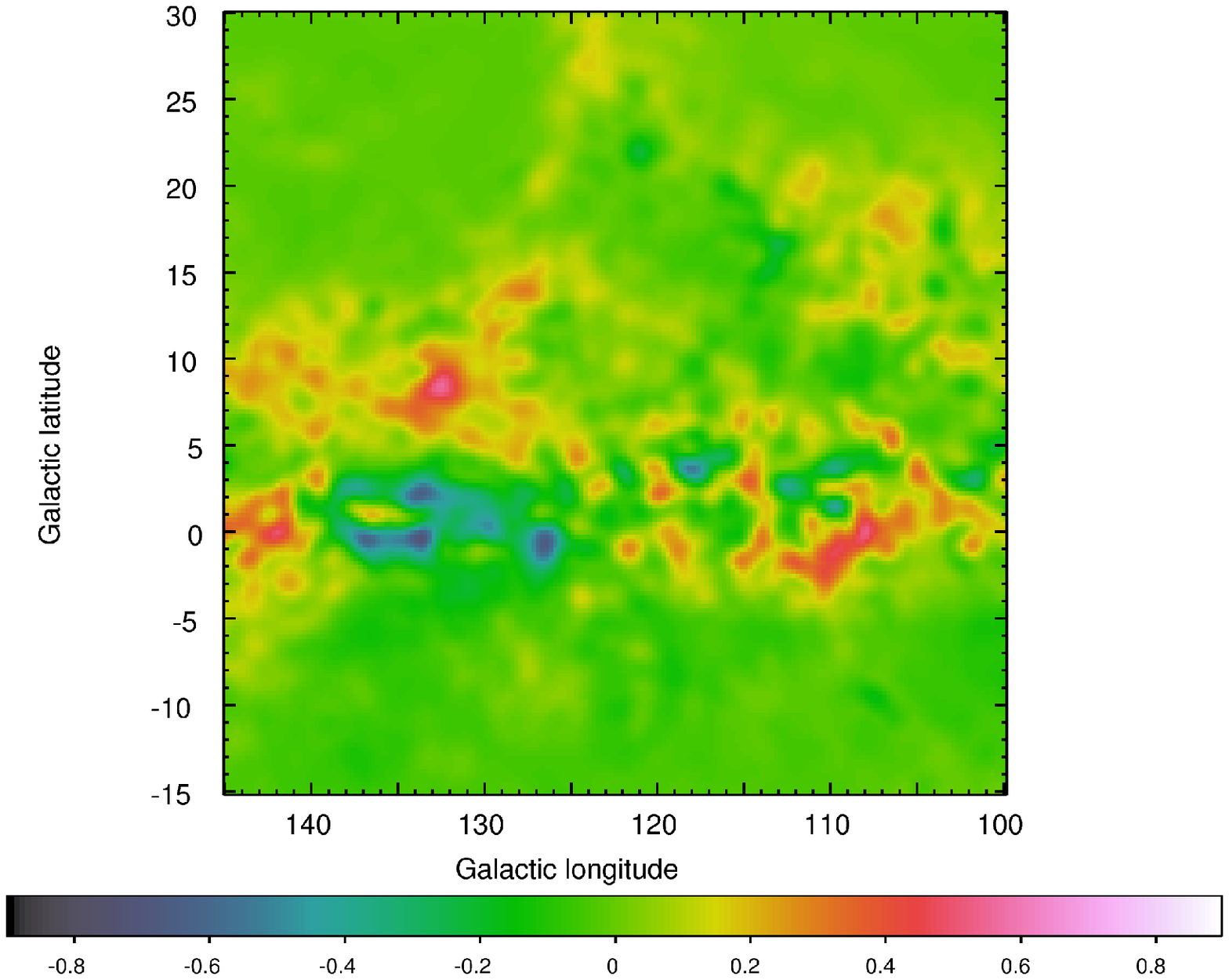}&
\includegraphics[width=0.5\textwidth]{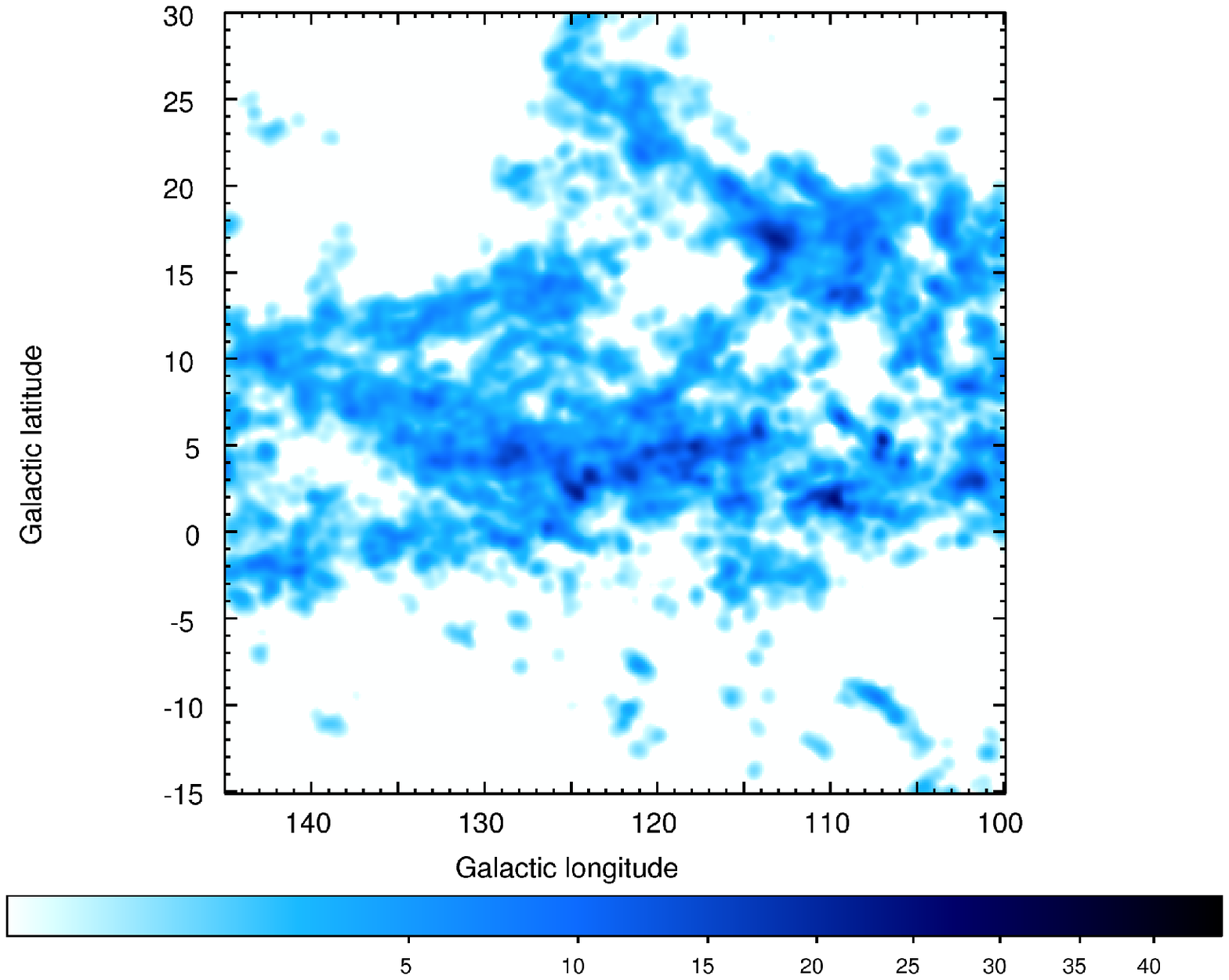}\\
  \end{tabular}
\caption{Maps of the region of Cassiopeia and Cepheus. Left: $\ebv$ residuals
(magnitudes), obtained from the map by Schlegel et al.~\cite{schlegel1998} after
the subtraction of the parts correlated with $\nhi$ and $\wco$ maps
\cite{cascep}. Right: $\wco$ (K km s$^{-1}$) in the Gould Belt
region, from the 2.6 mm survey of the CfA telescope
\cite{dame2001}.}\label{maps}
 \end{center}
\end{figure*}

Following the method proposed by Grenier et al.~\cite{grenier2005}, we have
used interstellar
dust as an additional gas
tracer under the assumption that dust grains are well mixed with gas in the cold
and warm phases of the ISM under study. We have used the total 
dust column densities provided by the color
excess $\ebv$ map by Schlegel et al.~\cite{schlegel1998}. From this map we have 
subtracted the parts linearly correlated with the best-fit combination of $\nhi$ and $\wco$
maps built for the four regions separated along the lines of sight, from the Gould Belt to the outer arm \cite{cascep}. 

The residual map is shown in Fig.~\ref{maps} on the left.
Within a few degrees from the Galactic plane, the map exhibits clumps of positive and negative
residuals. Their origin is unclear because of the intrinsic
limitations of the $\ebv$ map near the plane  \cite{schlegel1998}. At
intermediate latitudes
($|b|>5^\circ$), the map is dominated by structured envelopes of
positive residuals surrounding the Gould Belt CO clouds. For comparison the
$\wco$ map is shown on the right of Fig.~\ref{maps}. Smaller negative
residuals are found toward the dense CO cores.

\begin{figure}[!hbt]
\includegraphics[width=0.5\textwidth]{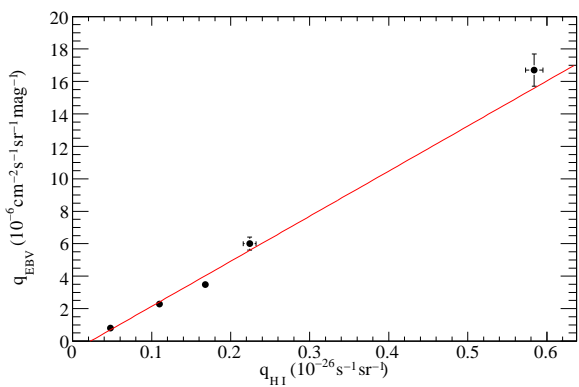}
\caption{Correlation between $\qebv$ and
$\qhinull$ in the Gould Belt over 5 energy ranges between 200 MeV and 10 GeV.
The solid red line represents the best linear fit.}\label{qhiebv}
\end{figure}

Adding the $\ebv$ residual map significantly improves the fit to the
\g-ray data~\cite{cascep}. 
Given the spatial correlation, we compared the \g-ray
emissivities per unit of
$\ebv$ residuals ($\qebv$) with the emissivities per $\hi$ atom ($\qhinull$)
in the Gould-Belt clouds. The results are shown in Fig.~\ref{qhiebv}.

The good linear correlation between $\qebv$ and $\qhinull$ over 2 decades
in energy suggests that the same process produces the \g{} rays associated with
$\hi$ and the dust-traced component.

The \g-ray emission
arising from hadronic interactions of CRs with dust grains is too faint to be
detected. The IC emission coming from dust infrared emission up
scattered by CR electrons is also expected to be faint and spectrally different
from the $\pi^0$-decay radiation associated with $\hi$ \cite{grenier2005}.
Therefore, the results point to
the presence of additional gas which is not properly accounted for in the $\nhi$
and $\wco$ maps. The negative residuals seen towards the CO cores can be
interpreted as small local variations in the dust-to-gas ratio or in the dust
thermal emissivity in the denser, well-shielded parts of the molecular
clouds. The nature of the additional gas, however, is not determined yet.

Substantial uncertainties result from the approximations applied to
handle the radiative transport of the $\hi$ lines. The
choice of a uniform spin temperature,
$T_S=125$ K, has often been adopted in the past, as in our study. Recent
absorption surveys
\cite{dickey2009} propose higher values in the outer Galaxy ($250 - 400$ K)
which
would yield $\nhi$ densities lower than the present estimates. Conversely,
colder and denser $\hi$ gas with $T_S$ as low as $30-40$ K is quite abundant in
the mid-latitude clouds \cite{heiles2003}. This phase contributes much larger
$\nhi$ column-densities than the estimates based on a uniform spin temperature
of a few hundreds K. Self absorption could also bias the $\nhi$ column
densities to low values.

On the other hand, at the interface between the atomic and CO-traced phase of
an interstellar complex we might find an additional phase where $\hd$ exists
but it is not well mixed with CO (e.g. because $\hd$ is more efficiently
self-shielded than CO in the low extinction range) or densities are too low to
excite the CO 2.6 mm line transition \cite{lombardi2006}. A poor correlation
with $\wco$ maps and, conversely, a good correlation with the $\ebv$ color
excess has
already been shown for alternative molecular tracers (like the 9 cm
line of CH) in translucent molecular clouds, and possibly proposed for the
translucent envelopes of giant molecular clouds \cite{magnani2003}.

\bigskip

\begin{acknowledgments}
The \textit{Fermi} LAT Collaboration acknowledges support from a number of
agencies and institutes for both development and the operation of the LAT as
well as scientific data analysis. These include NASA and DOE in the United
States, CEA/Irfu and IN2P3/CNRS in France, ASI and INFN in Italy, MEXT, KEK, and
JAXA in Japan, and the K.~A.~Wallenberg Foundation, the Swedish Research Council
and the National Space Board in Sweden. Additional support from INAF in Italy
and CNES in France for science analysis during the operations phase is also
gratefully acknowledged.

LT is partially supported by the International Doctorate
on AstroParticle Physics (IDAPP) program.
\end{acknowledgments}


\bigskip

\begin{thebibliography}{99}

\bibitem{cascep}
A.~A. {Abdo} et~al., \apj \textbf{710}, 133 (2010),
  arXiv:0912.3618.

\bibitem{strong1998}
A.~W. {Strong} and I.~V. {Moskalenko}, \apj \textbf{509}, 212 (1998),
  arXiv:astro-ph/9807150.

\bibitem{strong2007}
A.~W. {Strong}, I.~V. {Moskalenko}, and V.~S. {Ptuskin}, Annual Review of
  Nuclear and Particle Science \textbf{57}, 285 (2007),
  arXiv:astro-ph/0701517.

\bibitem{lochiemiss}
A.~A. {Abdo} et~al., \apj \textbf{703}, 1249 (2009), arXiv:0908.1171.

\bibitem{dermer1986a}
C.~D. {Dermer}, \apj \textbf{307}, 47 (1986).

\bibitem{dermer1986b}
C.~D. {Dermer}, \aap \textbf{157}, 223 (1986).

\bibitem{mori2009}
M.~{Mori}, \astropp \textbf{31}, 341 (2009), arXiv:0903.3260.

\bibitem{hunter1997}
S.~D. {Hunter} et~al., \apj \textbf{481}, 205 (1997).

\bibitem{strong2004}
A.~W. {Strong}, I.~V. {Moskalenko}, and O.~{Reimer}, \apj \textbf{613}, 962
  (2004), arXiv:astro-ph/0406254.

\bibitem{nongevexc}
A.~A. {Abdo} et~al., \prl \textbf{103}, 25, 251101 (2009), arXiv:0912.0973.

\bibitem{digel2001}
S.~W. {Digel} et~al., \apj \textbf{555}, 12 (2001).

\bibitem{vela1}
A.~A. {Abdo} et~al., \apj \textbf{696}, 1084 (2009), arXiv:0812.2960.

\bibitem{solomon1991}
P.~M. {Solomon} and J.~W. {Barrett}, in \emph{Dynamics of Galaxies and Their
  Molecular Cloud Distributions}, edited by {F.~Combes \& F.~Casoli}, vol. 146
  of \emph{IAU Symposium}, p. 235 (1991).

\bibitem{digel1990}
S.~{Digel}, P.~{Thaddeus}, and J.~{Bally}, \apjl \textbf{357}, L29 (1990).

\bibitem{sodroski1995}
T.~J. {Sodroski} et~al., \apj \textbf{452}, 262 (1995).

\bibitem{israel1997}
F.~P. {Israel}, \aap \textbf{328}, 471 (1997),
arXiv:astro-ph/9709194.

\bibitem{strong2004grad}
A.~W. {Strong} et~al., \aap \textbf{422}, L47 (2004),
  arXiv:astro-ph/0405275.

\bibitem{digel1996}
S.~W. {Digel} et~al., \apj \textbf{463}, 609 (1996).

\bibitem{nakanishi2006}
H.~{Nakanishi} and Y.~{Sofue}, \pasj \textbf{58}, 847 (2006),
  arXiv:astro-ph/061076.

\bibitem{lin1996}
Y.~C. {Lin} et~al., \apjs \textbf{105}, 331 (1996).

\bibitem{albert2007ic443}
J.~{Albert} et~al., \apjl \textbf{664}, L87 (2007), arXiv:0705.3119.

\bibitem{aharonian2008w28}
F.~{Aharonian} et~al., \aap \textbf{481}, 401 (2008), arXiv:0801.3555.

\bibitem{acciari2009ic443}
V.~A. {Acciari} et~al., \apjl \textbf{698}, L133 (2009), arXiv:0905.3291.

\bibitem{lat2009w51}
A.~A. {Abdo} et~al., \apjl \textbf{706}, L1 (2009).

\bibitem{case1998}
G.~L. {Case} and D.~{Bhattacharya}, \apj \textbf{504}, 761 (1998),
  arXiv:astro-ph/9807162.

\bibitem{strong1988}
A.~W. {Strong} et~al., \aap \textbf{207}, 1 (1988).

\bibitem{lorimer2004}
D.~R. {Lorimer}, in \emph{35th COSPAR Scientific Assembly}, vol.~35 of
  \emph{COSPAR, Plenary Meeting}, p. 1321 (2004).

\bibitem{strong1996}
A.~W. {Strong} and J.~R. {Mattox}, \aap \textbf{308}, L21 (1996).

\bibitem{gibson2005}
S.~J. {Gibson} et~al., \apj \textbf{626}, 195 (2005),
  arXiv:astro-ph/0503117.

\bibitem{evoli2008}
C.~{Evoli} et~al., Journal of Cosmology and Astro-Particle Physics
\textbf{10},
  18 (2008), arXiv:0807.4730.

\bibitem{grenier2005}
I.~A. {Grenier}, J.-M. {Casandjian}, and R.~{Terrier}, Science \textbf{307},
  1292 (2005).

\bibitem{schlegel1998}
D.~J. {Schlegel}, D.~P. {Finkbeiner}, and M.~{Davis}, \apj \textbf{500}, 525
  (1998), arXiv:astro-ph/9710327.

\bibitem{dame2001}
T.~M. {Dame}, D.~{Hartmann}, and P.~{Thaddeus}, \apj \textbf{547}, 792 (2001),
  arXiv:astro-ph/0009217.

\bibitem{dickey2009}
J.~M. {Dickey} et~al., \apj \textbf{693}, 1250 (2009), arXiv:0901.0968.

\bibitem{heiles2003}
C.~{Heiles} and T.~H. {Troland}, \apj \textbf{586}, 1067 (2003),
  arXiv:astro-ph/0207105.

\bibitem{lombardi2006}
M.~{Lombardi}, J.~{Alves}, and C.~J. {Lada}, \aap \textbf{454}, 781 (2006),
  arXiv:astro-ph/0606670.

\bibitem{magnani2003}
L.~{Magnani} et~al., \apj \textbf{586}, 1111 (2003).

\end{thebibliography}
\end{document}